\documentclass[12pt]{iopart}

\usepackage{iopams}
\usepackage{bm}          
\usepackage[dvips]{graphicx}
\usepackage{float}
\usepackage{gensymb}
\def \totder #1#2 { \frac{ {\mathrm d} #1 }{ \mathrm d #2} }

\begin{document}

\title[Multiscale modeling of submonolayer growth]{Multiscale modeling of submonolayer growth for Fe/Mo(110)}

\author{M Ma\v{s}\'{\i}n$^1$, M Kotrla$^1$, B Yang$^2$, M Asta$^3$, M O Jahma$^4$ \\ and T Ala-Nissila$^{4,5}$}

\address{$^1$Institute of Physics, Academy of Science of the Czech Republic, Na Slovance 2, Prague, 18221, Czech Republic}
\address{$^2$Department of Chemical Engineering and Materials Science, University of California, Davis, CA 95616, USA.}
\address{$^3$Department of Materials Science and Engineering University of California Berkeley, Berkeley, California 94720, USA.}
\address{$^4$COMP CoE at the Department of Applied Physics, Aalto University School of Science,  P.O. Box 11000, FI-00076 Aalto, Espoo, Finland}
\address{$^5$Department of Physics, Brown University, Providence RI 02912-1843}

\ead{kotrla@fzu.cz}

\begin{abstract}
We use a multiscale approach to study a lattice-gas model of
submonolayer growth of Fe/Mo(110) by Molecular Beam Epitaxy. To begin
with, we construct a two-dimensional lattice-gas model of the
Fe/Mo(110) system based on first-principles calculations of the
monomer diffusion barrier and adatom-adatom interactions. The model is
investigated by equilibrium Monte Carlo (MC) simulations to compute
the diffusion coefficients of Fe islands of different sizes. These
quantities are then used as input to the coarse-grained Kinetic Rate
Equation (KRE) approach, which provides time evolution of the island
size distributions for a system undergoing diffusion driven
aggregation within the 2D submonolayer regime. We calculate these
distributions at temperatures $T=500$ K and $1000$ K using the KRE
method.  We also employ direct kinetic MC simulations of our model to
study island growth at $T=500$ K, and find good agreement with the KRE
results, which validates our multi-scale approach.
\end{abstract}

%


\maketitle

\section{Introduction}
\label{sec:intro}
Basic concepts of growth of thin films were already developed in the last century, in particular in
the context of the classical nucleation theory (for a review see e.g. \cite{venables1984nucleation,reichelt1988}).
Most recently, increased effort has been devoted to
understanding surface growth at the microscopic level both
experimentally and theoretically (for review see
e.g. \cite{michely2004islands,kotrla1996numerical} and references
therein). Due to the inherently complicated nature of surface
growth processes numerical simulation methods at atomic scales, such as
kinetic Monte Carlo (KMC) and Molecular Dynamics (MD) methods,
have played an important role in modeling surface growth.  However,
realistic modeling of surface growth in metallic systems under
Molecular Beam Epitaxy (MBE) conditions is a challenging
problem, which in addition to atomic scale simulation methods
benefits greatly from development of efficient multi-scale
models. One of the main challenges in trying to model MBE growth processes at
atomic scales stems from the large difference between the deposition
and diffusion time scales under typical MBE conditions.  In addition, classical
activation barriers controlling atomic
diffusion mechanisms on growing surfaces may span a wide range of values, from
meV to eV scales. This makes it
impractical to apply classical MD simulations without
using flux values that are orders of magnitude larger than those in
MBE. This naturally raises serious doubts concerning the validity of
such simulations to faithfully describe MBE.

To this end, atomistic calculations (using either {\it ab initio} or classical
semi-empirical potentials) have been frequently employed to construct
solid-on-solid (SOS) type of interaction Hamiltonians, which can then be
used in kinetic Monte Carlo simulations. Such KMC simulations are
applicable to scales from hundreds to a few thousands of atomic
distances. Although significant progress has been recently made in
speeding up KMC simulations
\cite{jonsson11,nandipati10PhysRevB.81.235415} they remain
computationally demanding in most realistic cases. Thus, appropriately
coarse-grained methods are beneficial. One such promising approach is
based on describing the kinetics of growth with the classical kinetic rate
equation (KRE) approach
\cite{Stoldt1999,PhysRevE.58.4037,koponen:086103,Evans2006}.
It has been shown by extensive studies that the KRE approach can
accurately describe 2D growth processes in cases where spatial
correlations do not play an important role. The main advantage of the
KRE approach over atomistic simulation methods is that the numerical
effort required is relatively insensitive to the ratio between the deposition
flux and adatom diffusion scales, and thus realistic growth parameter
values can be easily studied (including fluxes much less than $1$ ML/s).
The conceptual simplicity of the KRE theory also allows easy adjustment
of the various growth processes and their rates.  Most recently, an
improved self-consistent KRE approach to irreversible island growth has
been developed by Hubartt {\it et al.} \cite{Hubartt2011}, which
accurately reproduces results obtained from KMC simulations.

As far as specific systems are concerned there exist a number of
experimental studies with atomistic resolution for different metals on
bcc(110) surfaces, which are of great current interest because many of
the important magnetic metals have bcc crystal structure. For example,
Fe on Fe(110) \cite{koehler2000scanning,koehler2000investigation}, Fe
on W(110) or Mo(110) \cite{fruchart2007growth}, and Cu on W(110)
\cite{reshoef1999atomistics} have been studied. Modeling and
understanding of growth on this anisotropic surface is still very
limited. In the case of homoepitaxy, there are atomistic KMC
simulations of Fe/Fe(110) homoepitaxy
\cite{koehler2000scanning,koehler2000investigation}, and simulations
of a niobium film by molecular dynamics \cite{marchenko2007computer}.
On a larger length scale there is a qualitative phase-field simulation
of stripe arrays observed in multilayer growth on a bcc(110) surface
\cite{yu2008phase}.  In the case of heteroepitaxy, modeling becomes
even more complex because in addition to anisotropy one should take
into account also strain (lattice mismatch) effects. As regards
atomistic modeling of bcc(110) surfaces there are, to our knowledge,
only two published KMC
simulations: investigation of low temperature MBE growth of Cu on
W(110) \cite{rogowska2002anisotropic} and study of island morphology
in the Fe/Mo(110) system \cite{goykolov2009study}. These simulations
use lattice gas models with effective parameters for the
heteroepitaxial system. Therefore, the effect of strain is included
only indirectly.  Another deficit in these studies is that adatom
diffusion, which plays a predominant role in MBE growth, is not
described in terms of realistic energy parameters. The situation for
heteroepitaxy on bcc(110) surfaces is in this sense unsatisfactory so
far. Contrary to many calculations of diffusion barriers on fcc
surfaces, e.g.  in the case of well studied Pt/P(111) system
\cite{boisvert98}, there are only a few results for the quantitative
values of diffusion barriers on bcc(110) surfaces
\cite{gollisch1986adsorption}.

In this work, we present a comprehensive multiscale study of
heteroepitaxial growth on a bcc(110) surface in the submonolayer
regime. Our aim is to present a multiscale modeling strategy using
data obtained from microscopic {\it ab initio} electronic structure
calculations as a basis to construct models, which allow numerical
studies of MBE growth.  As a specific system, we have chosen Fe
heteroepitaxy on Mo(110) due to its interesting applications to
magnetic nanoislands
\cite{malzbender1998epitaxial,jubert2001selfassembled}.
The lattice mismatch between Fe and Mo is relatively
large ($- 9$ \%) and thus we restrict
the modeling here to early stages of growth, {\it i.e.} to the
submonolayer regime of 2D island growth, which is dominated
by deposition and surface diffusion of Fe adatoms.
It is know that strain influences surface diffusion \cite{brune1995effect}
in such a way that
tensile strain in general leads to a decrease of diffusion barriers
\cite{schroeder1997diffusion,sabiryanov2003growth}.
In our case this effect is included in the first-principles
energy barriers obtained for Fe diffusion. In this work no
other strain effects are included.
Using the {\it ab initio} data we then
construct a lattice-gas Hamiltonian with correct energetics, and use
this Hamiltonian in KMC simulations.  Further, we employ equilibrium
MC simulations to calculate diffusion coefficients of islands of
various sizes. With these data we construct a KRE aggregation model,
which can then be used to study island size distribution development
on different scales and in different growth regimes.  We compare
results obtained from KMC and KRE approaches and find good agreement
between the two methods.

\section{Model and quantities calculated}
 \label{sec:model}

\subsection{Submonolayer model for Fe/Mo(110)}
 \label{subsec:subML_model}

\begin{figure}
\centering
\includegraphics[width=100mm]{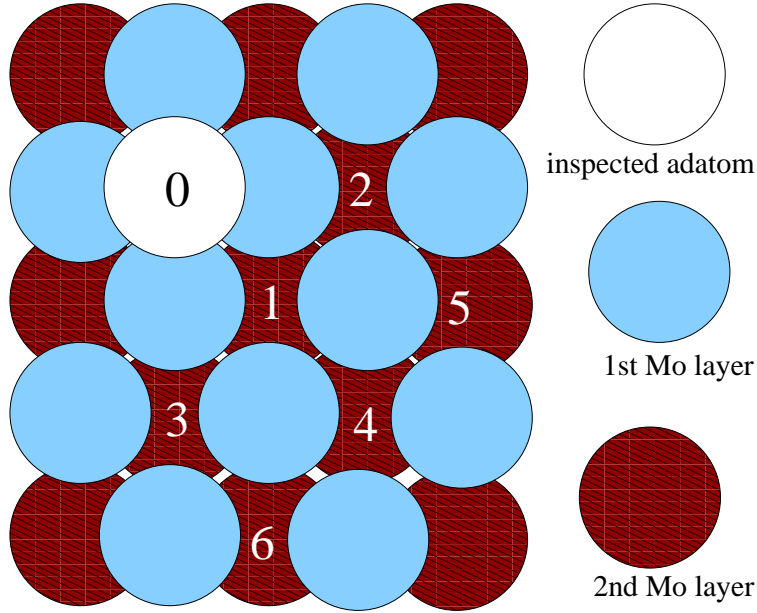}
\caption{(Colour online) A schematic of an ideal Mo(110) surface,
with one Fe adatom occupying site 0, which is an on-top site
for the {\it second} layer Mo atoms. A second
Fe adatom that forms a Fe-Fe dimer
can then occupy different neighboring positions,
such as 1 (nearest neighbor), 2 (second nearest neighbor), and
so on up to 6th nearest neighbors in the present case.}

\label{FIGURE_neighbors}
\end{figure}

We begin by constructing a growth model of the Fe/Mo(110) surface,
which will be used to study diffusion and growth of Fe adatoms on an ideal Mo(110)
surface. It is based on the well-established solid-on-solid (SOS) model
of epitaxial growth \cite{clarke88-JApplPhys63.2272} that has been shown to be
capable of reproducing many aspects of epitaxial growth, including
scaling of island densities
\cite{ratch94-PhysRevLett.72.3194}.
In the SOS approach neither vacancies nor overhangs are permitted, and the substrate is
assumed to have the bcc(110) structure with no steps. In the present work we apply this model
in the range of submonolayer coverage, which means that it corresponds to a 2D lattice-gas model, where
diffusion is restricted to the 2D surface and particle deposition occurs on empty lattice sites only.
We consider lateral interactions up to sixth-nearest neighbors
as shown schematically in figure \ref{FIGURE_neighbors}.

To construct the atomistic model, we first need to determine the relevant Fe-Mo and Fe-Fe
adatom interactions on the Mo(110) surface.
It is experimentally known \cite{murphy02PhysRevB.66.195417}
that a monolayer of Fe deposited on Mo(110) forms a commensurate
layer and thus we take the adsorption sites of the Fe adatoms to be on top of the second atomic layer of the
Mo(110) surface,
as shown in figure 1.
The interaction with the substrate is
characterized by a hopping barrier for diffusion of a free adatom,
described by an activation energy $E_d$. We consider Fe-Fe adatom interactions
up to sixth-nearest neighbors described by six binding energies $E_{Bk}$, $k = 1, ..., 6$.
To compute the activation energies and binding energies we performed
first-principles calculations within the framework of density-functional theory,
employing ultrasoft pseudopotentials \cite{Vanderbilt90-PhysRevB.41.7892},
as implemented in the Vienna ab initio
simulation package (VASP) \cite{Kresse96a-PhysRevB.54.11169,Kresse96b}.
Calculations were performed employing the 1991
generalized gradient approximation of Perdew and Wang \cite{Perdew92-PhysRevB.46.6671} and the
pseudopotentials labeled "Mo" and "Fe" in the VASP library.

The activation
energies and binding energies reported here were obtained from spin-polarized
calculations employing slab supercells containing seven Mo layers with lateral
periodic lengths corresponding to $4 \times 4$ primitive unit cells on the Mo(110)
surface.  In the calculations Fe adatoms were arranged in appropriate
configurations on top and bottom surfaces.  A plane-wave cutoff of $300$ eV and
$k$-point mesh of $4 \times 4 \times 1$ were employed in all calculations.  A series of
calculations was undertaken to check the convergence of the results with respect
to plane-wave cutoff, slab and vacuum thickness, and the lateral periodic
dimensions of the supercell.  Based on these calculations the activation energy is
estimated to be converged numerically to within a few times $0.01$ eV.
Using the above procedure, we obtained the following values of the model parameters:
$E_d = 0.41$ eV,
$E_{B1} = 329$ meV, $E_{B2} = 72$ meV, $E_{B3} = 79$ meV,
$E_{B4} = - 10$ meV, $E_{B5} = - 4.1$ meV, and $E_{B6} = 7.5$
meV. In the present 2D case it can be immediately concluded that
the last three values are probably small enough to be omitted in the simulations.
We have verified this by testing their influence, which will be discussed in section
\ref{subsec:diff_coeff}.

\subsection{Calculation of adatom island diffusion coefficients}
\label{subsec:diff_coeff}

As the first step in the construction of a coarse-grained growth model based on the KRE approach we
need to determine the tracer diffusion coefficients
$D_{s}(T)$ of islands of size $s$. In this work, they are determined using equilibrium Monte Carlo
simulations.  Based on the previous section, the corresponding 2D lattice-gas Hamiltonian reads
\begin{equation}
H = - \sum_{k=1}^{6} \sum_{\vert i-j \vert = k, i > j } E_{Bk} m_{i} m_{j} ,
\label{hamiltonian}
\end{equation}
which includes pair interactions  $E_{Bk}$ only up to sixth neighbors. Here,
$m_{i} = 0,1$ is the occupation number of a lattice site $i$. We note that equation (1)
corresponds to simple bond counting between each pair of Fe adatoms.

Diffusion is modeled by thermally activated hops of Fe adatoms to a
vacant nearest neighbor position. In this section, our model is studied by
standard Monte Carlo simulations with Metropolis dynamics using the
transition dynamics algorithm (TDA), which is a good choice for
calculation of diffusion coefficients and is described in detail in
Refs. \cite{Ala-Nissila92-PhysRevB.46.846,Vat98}.  The jump
probability in TDA depends on the energies in initial and final sites
and the saddle point energy.  The energy of the saddle point is
calculated as an arithmetic mean of the energies of the initial and
final sites, including the diffusion barrier for a free adatom, {\it
 i.e.}
\begin{equation}
E_{S} = \frac{1}{2}\left( E_{i} + E_{f}\right) + E_{d}.
\end{equation}
The TDA in our simulations
evaluates particle jumps in two successive steps. In the first
step, we evaluate the jump probability from the initial site to the saddle
point. This step is accepted with the probability $\nu \exp( -\Delta E/ kT)$, where
$\Delta E = E_{S} - E_{i}$, except for the case where the
energy of the initial site is higher than the energy of the saddle point,
in which case the jump is always accepted (it could occur e.g. in the case of
strong adatom-adatom repulsion). The process is repeated for the next step -- a move from the saddle
point to the final position. This step is accepted with the probability $\nu \exp( -\Delta E/ kT)$,
where $\Delta E = E_{f} - E_{S}$ except for the case when
the energy of the saddle point is higher than the energy of the final point.
This scheme is more realistic than adding
the isolated adatom barrier to half of the difference between
the final and initial site energies.

The tracer diffusion coefficient $D_{s}(T)$ is computed from the mean square displacement
of a virtual particle representing the center-of-mass of the whole island:
\begin{equation}
D_{s} = \lim_{t \rightarrow \infty} \frac{1}{2t}
           \left< \left| \vec{R}_{s}(t) -
           \vec{R}_{s}(0) \right|^2 \right>,
\label{tracer}
\end{equation}
where $ \vec{R}_{s}(t) = (1/s)\sum_{j=1}^s [\vec{r}_{j}(t) -
 \vec{r}_{j}(0)] $ represents the position vector of the
center-of-mass of the individual island, $s$ is number of particles in
the island, and $\vec{r}_{j}(t)$ is
the position of the particle $j$ in the island at the time $t$.  As
usual, we have computed $D_{s}(T)$ by using the memory expansion
method \cite{Ala02,Yin98}.  To determine $D_{s}(T)$ for a given island
of size $s$ we prepare an initial configuration composed of $s$
particles comprising one island.  Detachment of particles from the
island is explicitly forbidden.

\begin{figure}
\centering
\includegraphics[width=100mm]{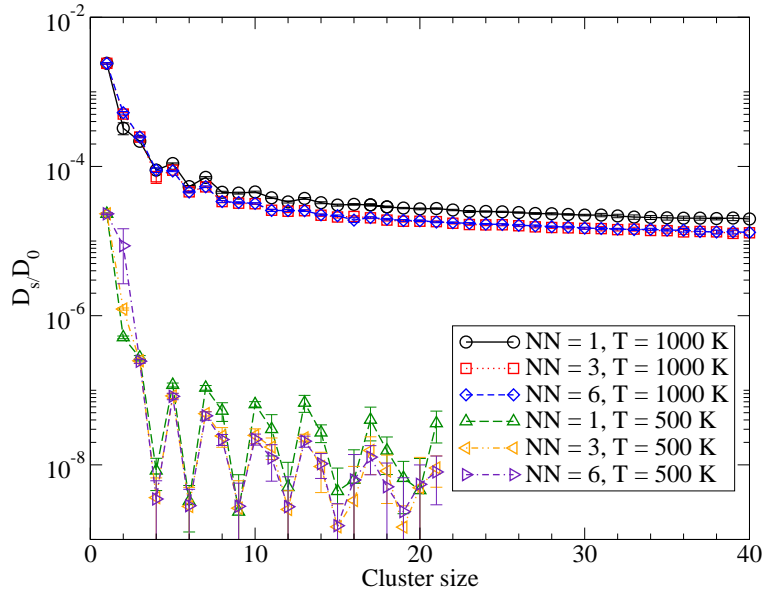}
\caption{(Colour online) Dependence of normalized diffusion
  coefficients on island size $s$ for the two temperatures considered
  here, with three different ranges of interactions, as specified by
  the label NN (see text for details).}
\label{FIGURE_diffusion}
\end{figure}

We have calculated $D_{s}(T)$ at two temperatures: $T = 500$ K and $T
= 1000$ K.  Results of our simulations are shown in
figure~\ref{FIGURE_diffusion} with data averaged over $20$ runs. We note
that all the values of
$D_s$ are normalized by the prefactor $D_0 = (1/4) \nu_{\mathrm osc}
a^2$, where $a=3.15$ \AA\, is the lattice constant of Mo and $\nu_{\mathrm
osc}$ is the vibrational frequency.
There are some clusters with metastable configurations, which cause
large error bars for some points in figure~\ref{FIGURE_diffusion} at the
lower temperature.  The diffusion coefficients of islands depend on
size non-trivially and can even oscillate as a function of size $s$
\cite{wan90,wan98,kyu00,hei99}.  These oscillations can be in part
understood by surface geometry: certain close-packed island
configurations can be more stable than
others \cite{tru00,sal01,tru01,tru01B}.  In our calculations, the
oscillations as a function of size $s$ are clearly seen at the lower
temperature.  As temperature increases, the oscillations are dampened,
and at our highest temperature the diffusion coefficient is rather
well described by $D(s) \sim s^{-3/2}$, which is expected if island
diffusion is limited by atomic motion along the perimeter of an
island \cite{kha95,kha96}.  For Fe/Mo(110), there is a strong
dependence of $D_{s}$ on temperature here because the (bare) single
adatom diffusion barrier is about $0.4$ eV.

To check the robustness of the diffusion coefficient against the range
of Fe-Fe interactions, we performed calculations for three different
cases: (i) with nearest neighbor interactions only, (ii) interactions
up to third nearest neighbors and (iii) full set of interactions up to
sixth nearest neighbors.  Figure \ref{FIGURE_diffusion} clearly
demonstrates that in the present model the main features of the size
dependence of the diffusion coefficient shows the same oscillations
for all three different cases studied here.  On a quantitative level
we can conclude that it is sufficient to include interactions only up
to third nearest neighbors when considering KMC simulations of the
growth model.  This significantly speeds up and simplifies the
simulations.

\subsection{Island size distribution}
 \label{subsec:distribution}

The fundamental quantity in 2D island growth is the size distribution
function, which characterizes the number density $n_s$ of islands of size $s$ at any
given time during growth. For simplified kinetic growth models this
function often shows scaling behaviour, which gives valuable
information about the growth process
\cite{Bartelt1992,bales1994,bales1995}. In general, however, there is
no theoretical guarantee for the existence such simple scaling
solutions. This is true in particular when realistic size-dependent
diffusion coefficients are used \cite{jahma2005}. However, there can
still exist stationary phases of growth, where the effective scaling
laws hold in a limited region of parameter space
\cite{KoponenJahmaRusanenAlanissila_PRL2004}.

To this end, in the present work we consider scaled distribution
functions, which can be obtained by defining a probability density
that a randomly selected atom is contained in an island of size $s$,
viz.\ $p_s(\theta)=s n_s(\theta)/\theta$, where $\theta$ is the
surface coverage and $n_s$ is the island number density of islands of
size $s$, {\it i.e.} the number of islands of size $s$ divided by the number
of occupation sites. The average island size is defined as
$\bar{s}=M_2/M_1$, where the $k^{\mathrm{th}}$ moment of the
distribution is given by $M_k=\sum_1^\infty s^k n_s(\theta)$. The
scaling function is defined as $g(x)=\bar{s} p_s(\theta)$, where
$x=s/\bar{s}$\, \cite{KoponenJahmaRusanenAlanissila_PRL2004}. In the
scaling regime this function becomes independent of coverage for a
given set of relevant growth parameters.

\section{Numerical studies of island growth}
\label{sec:procedures}
\subsection{Atomistic modeling of MBE growth by KMC}
 \label{subsec:KMC_growth}

The most direct way to study MBE growth of the Fe/Mo(110) system
is to employ KMC simulations using the lattice-gas approach.
Diffusion is modeled by thermally activated hops of adatoms to unoccupied
nearest neighbor positions with a configuration dependent rate.
The probability $P_{i}$ of a jump of an adatom from a site $i$
is exponentially proportional to the product of an activation energy
$E_{\mathrm{A}(i)}$ and the inverse temperature $1/T$
\footnote{We note that instead of the TDA algorithm, which we employed for
the calculation of the equilibrium island diffusion coefficients, we have here chosen
the commonly used simple exponential function to enhance the numerical efficiency of the KMC
growth simulations.}
\begin{equation}
P_{i} = \nu_d \exp \left(-\frac{E_{\mathrm{A}(i)}}{k_B T}\right),
\end{equation}
where the prefactor $\nu_d
  = \nu_s\nu_{osc} = 1.6 \times 10^{13}$ $s^{-1}$, and $\nu_s=4$ is the number
 of equivalent saddle points and $\nu_{osc}=4\times10^{12}$ is
 the oscillation frequency.  The activation energy $E_{\mathrm{A}(i)}$
for the chosen adatom is given by sum of the activation energy for a
free particle $E_d$ and the added interactions from neighboring
occupied sites.  Here, we use the standard bound-counting scheme
according to the lattice-gas description. As justified in section II B,
we include Fe-Fe interactions up to the third nearest neighbors
only. Hence,
\begin{equation}
E_A(i) = E_d + l_1 E_{B1} + l_2 E_{B2} + l_3 E_{B3},
\end{equation}
where $E_{B1}$,  $E_{B2}$, and $E_{B3}$ are interaction energies of the
adatom with its lateral first, second and third nearest neighbors, correspondingly, obtained
by first-principles calculations in section  \ref{subsec:subML_model}.
The integer coefficients $l_1$, $l_2$, and $l_3$ count the number of occupied first, second, and third
nearest neighbors of the site $i$, correspondingly, which means that
$l_1 \in [0,4]$, $l_2 \in [0,2]$, and $l_3 \in [0,2]$.

In our KMC simulations the growth is reversible and it includes process of detachment of an
adatom from an island, and breaking of an island.
Our KMC simulations are thus consistent with the classical nucleation theory.
We note that in our KMC approach
islands diffuse by single adatom moves only; no collective adatom events have been
implemented here. To maximize the efficiency of the KMC method we utilized
the Bortz-Kalos-Lebowitz (BKL) algorithm with the implementation by Maksym
\cite{maksym88}.
Results reported in the next Section
represent simulations on a lattice of the size ranging from $500 \times 500$ to
$1000 \times 1000$ lattice sites, the number of runs varied from $20$ up to $1000$.

\subsection{Kinetic rate equation model of growth}
\label{subsec:rate_equation}

Despite the application of the BKL algorithm to speed up KMC
simulations, the microscopic approach is not feasible for large-scale
studies of island growth in the micrometer scale. This is mainly due to the fact that in real
MBE growth under typical conditions the (dimensionless) ratio $\kappa
= \Phi a/D_1 \ll 1$.  Here $\Phi$ is the incoming adatom flux, $a$ is
the lattice constant (microscopic length) and $D_1$ is the
single-particle diffusion coefficient. In real units, typical fluxes
under MBE conditions are a few monolayers per second, while diffusion
coefficients on metal surfaces typically have activation barriers less
than 1 eV (here $E_d = 0.4$ eV), leading to $\kappa \ll 1$. Simulating
such conditions with KMC is challenging, although most recent work shows that
fluxes as small as $10^{-3}$ are feasible \cite{Shim2012}.

To facilitate large-scale studies of MBE growth with small values
of $\kappa$, we employ the coarse-grained kinetic rate equation (KRE)
method \cite{PhysRevE.58.4037,koponen:086103}.  The rate equation
giving the time evolution of island size distribution $n_s$ for a
system undergoing diffusion driven aggregation and with incoming flux
$\Phi$ of adatoms per monolayer is given by
\begin{equation}
\totder{n_s(t)}{t} = \sum_{i+j=s} K(i,j) n_i n_j - \sum_{i=1}^{s-1} K(i,s)
n_i n_s + \Phi\delta_{1s}.
\label{RE}
\end{equation}
The function $K(i,j)$ is called aggregation kernel and is given by the usual ansatz
\begin{equation}
K(i,j)=D_i+D_j,
\end{equation}
where $D_i$ is diffusion coefficient of an adatom island of size $i$ on the
surface. In the present model, we neglect
adatom detachment from the islands, and explicit island capture rates
\cite{Hubartt2011}. In a previous study of submonolayer growth in the
Cu/Cu(100) and Cu/Cu(111) systems \cite{jahma2005}, the island diffusion
coefficients were calculated using KMC simulations and then used to calculate the corresponding
island size distributions using the KRE approach in the case of hyperthermal deposition. Similar
strategy is adapted here for normal MBE growth; however, unlike in reference \cite{jahma2005}
our energy parameters and thus the diffusion coefficients in figure 2 are based on {\it ab initio} data.

The coupled set of ordinary differential equations constituting the KRE model give the time evolution
of the island size distribution on the surface. It may be solved directly by
numerical integration, but given appropriate reaction rates this set of
differential equations is usually stiff, thus making the direct
approach demanding. In addition, obtaining complete analytic
solution is possible only in a limited number of cases with suitable rate
coefficients without making strong assumptions about the form
of the scaling function. Therefore, we solve these equations using the particle
coalescence (PCM)
method
\cite{KangRedner_PRL1984,KrapivskyMendesRedner_EPJB1998,KoponenJahmaRusanenAlanissila_PRL2004}.
In PCM a point-island approximation is used,
which is valid at low coverage or large island separations.
Aggregation events occur with probabilities specified by
the corresponding reaction kernels, and the deposition with the rate
proportional to the given adatom flux. An event is then randomly chosen
with a probability weighted by the corresponding rate. Since KREs describe
the system in the mean-field limit, there is no information on spatial
correlations in the system. The main advantage of the KRE approach over KMC
simulations, however, is that a wide range of values of $\kappa$ can be studied, in
particular the limit where $\kappa \ll 1$, because
all atomistic events in island dynamics are embedded into the effective rates.

\label{sec:scaling}
In a previous study \cite{koponen:086103} the island size distributions were found to be of
scaling form and the mean island size had a power-law form.
Even if the rates were not a homogeneous functions of
the island size, well-defined effective scaling
exponents for the mean island size and the size distribution function could be determined.
In the present case, we expect that scaling is invalidated due to
realistic diffusion coefficients which have a naturally inhomogeneous
form, as is evident in figure 2. There is no guarantee of either the existence of usual scaling type
of solutions or uniquely defined scaling exponents for the mean
island size and island densities. Thus, we have not tried to extract any
scaling exponents in the present work.

\section{Results and discussion}
\label{sec:results}
\begin{figure}
\centering
\includegraphics[width=100mm]{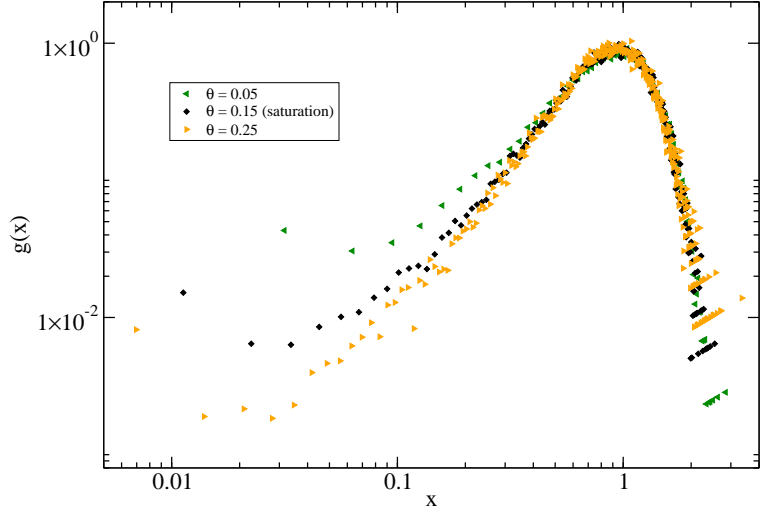}
\caption{(Colour online) Time development of scaled island size distribution in three
different coverage (deposition times), namely $\theta=0.05$,
$\theta=0.15$ (saturation coverage), and $\theta=0.25$ for $\Phi = 100$ ML/s
and $T=500$ K using the KMC approach.}
\label{time_devel}
\end{figure}

In this subsection, we consider  the scaled distribution
function $g(x)$, where $x= {\bar s}/s$. To start with, we have tested the
robustness of the growth model by using direct KMC simulations, as
described in the previous section. The BKL algorithm allows us to obtain
reasonable statistics at the temperature $T=500$ K if the flux is not too small.
We have first considered the case of a high flux of $\Phi = 100$ ML/s
(which corresponds to $\Phi/D_1=4.3 \times 10^{8}$ ML/cm$^{2}$)
and checked how the distribution varies with coverage.
Here data has been averaged over $20$ runs.
The results are shown in figure~\ref{time_devel}. As expected, for a fixed
coverage the function has a characteristic
form, with a strongly asymmetric peak around $x=1$. The long tail
extending to small island sizes reflects the high mobility of adatoms and small islands. Interestingly enough
we find that in all three cases (coverages 0.05, 0.15, and 0.25 ML)
the scaled function is almost identical except for the part where $x \ll 1$. The observed
decrease of $g(x)$ in the region of small $x$ with increasing coverage is an expected consequence of increased
aggregation to large islands. Based on the robustness of the scaled distribution we will only discuss results for the
saturation coverage of $\theta = 0.15$ in what follows.
We note that at $T=500$ K the Fe islands have a faceted, rhombic shape, as demonstrated in a
recent study using the present lattice--gas model \cite{goykolov2009study}.

\begin{figure}
\centering
\includegraphics[width=100mm]{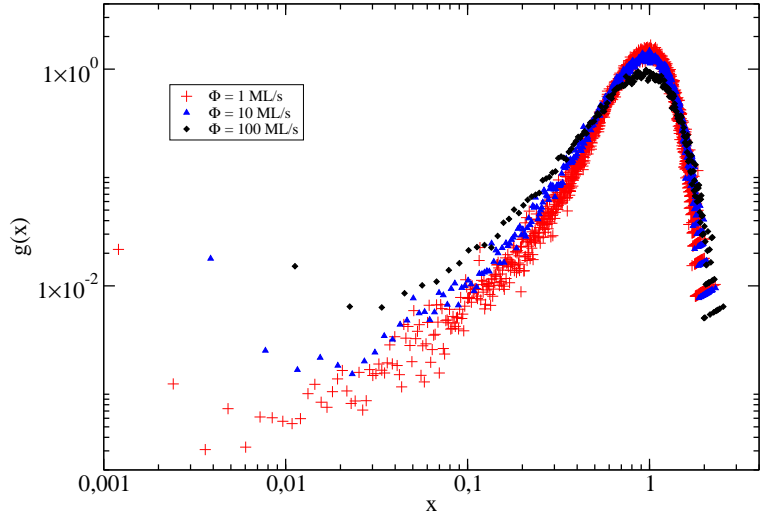}
\caption{(Colour online) Scaled island size distribution for incident
fluxes $\Phi=1, 10$ and $100$ ML/s from KMC simulations. In all
cases, data have been taken at the saturation coverage of $\theta = 0.15$.
See text for details.}
\label{flux_variation}
\end{figure}

At the next stage we studied the robustness of $g(x)$ against changing
the adatom flux at $T=500$ K.  We found that for the Fe/Mo system it
was very hard to acquire any meaningful KMC data for a flux value
below $\Phi = 1$ ML/s.  Nevertheless, for $\Phi = 1$ ML/s we were able
to obtain valid data after extensive averaging.  The results are shown
in figure~\ref{flux_variation}.  The data for $\Phi = 1$ ML/s represent
an average over $1000$ runs.  For comparison, we also calculated the
island distribution $\Phi = 10$ ML/s (averaged over $100$ runs) and
$\Phi = 100$ ML/s (averaged over $20$ runs).  Despite the poor
statistics it can be seen that the scaled distribution is relatively
insensitive to the flux, except for $x \ll 1$ where again there is
some deviation between the different fluxes. The reason behind this is
that even for relatively large fluxes, the fundamental processes
involved in island growth are the same and the character of the {\it
  scaled} island size distribution relatively insensitive to the flux.
However, the important conclusion here is that for the present model
the scaled distributions allow us to vary the flux by two orders of
magnitude from $\Phi = 1$ ML/s up to $100$ ML/s without any
significant change in the results.

At the next stage we employed the KRE approach to study island
growth. Using a flux of $100$ ML/s and at first keeping $T=500$ K
allows us to quantitatively compare the KRE results to those obtained
from KMC in figures 3 and 4. The advantage of the KRE approach over KMC
is evident in that the KRE data can be averaged over several hundreds
of runs here with relatively modest computational effort.  At this
stage it is also interesting to compare the KRE predictions in two
different growth modes. The KRE approach easily allows adjustment of
the various average rates in the kinetic equation. To this end, we
consider here two different growth modes. In the so-called limited
adatom-island mode, only adatoms are mobile while the islands are
immobile. This should be a reasonable approximation at low
temperatures (as compared to $E_d$) when the single adatom diffusion
coefficient $D_1$ is much larger than that of the dimers. As seen in
figure 2, this is not an unreasonable approximation at $T=500$ K. In the
second island-island mode, we solve the full KRE equations keeping all
the diffusion coefficients presented in figure 2.

\begin{figure}
\centering
\includegraphics[width=100mm]{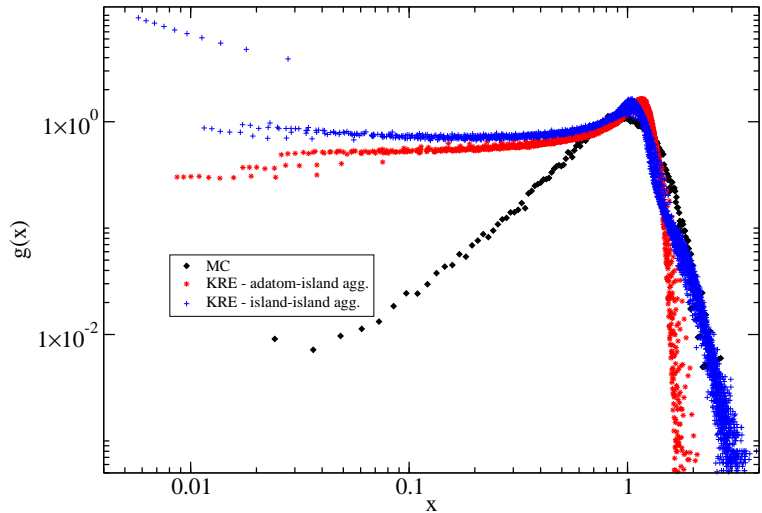}
\caption{(Colour online) Comparison between KMC simulations and KRE data in
the adatom-island and island-island growth modes at $T=500$ K.}
\label{T500}
\end{figure}

Comparison between the two different KRE growth modes and the KMC
results are shown in figure~\ref{T500}.  First, it can be observed that
even at this temperature the adatom-island aggregation mode causes
significant changes in the scaled distribution for larger island
sizes, which are strongly influenced by zero island mobility. In contrast,
the KMC and island-island mode KRE results are in excellent agreement
for $x > 5$.  There is, however, significant difference in $g(x)$ for
small island sizes.  This is in part due to the mean-field nature of
the KRE theory, which underestimates fluctuation-induced nucleation of
small islands, and in part due to the approximation of neglecting
adatom detachment from islands, which also enhances aggregation of
smaller islands to larger ones.

\begin{figure}
\centering
\includegraphics[width=100mm]{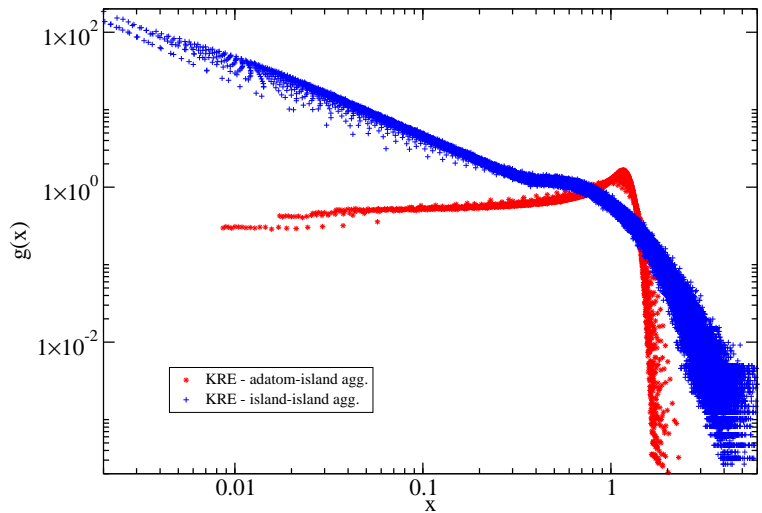}
\caption{(Colour online) Results of KRE for the adatom-island and island-island modes at
temperature $T=1000$ K. In this case, it was not possible to obtain KMC data.}
\label{T1000}
\end{figure}

Finally, in figure \ref{T1000} we show results from KRE for $T=1000$ K. In this case, large fluctuations and
finite-size effects prevented us from obtaining any statistically meaningful KMC data. It is interesting to note that
while the adatom-island mode data shows very little temperature dependence as compared to the case where
$T=500$ K, the distribution for the island-island mode is qualitatively different from that of figure \ref{T500}.
There is a significant enhancement of small islands in the regime $x \ll 1$, which indicates that in this case
neglecting the mobility of islands leads to significant errors in $g(x)$. This can be expected since figure 2 shows that
at $T=1000$ K island mobility is no longer orders of magnitude smaller than the single adatom mobility.

\section{Conclusions}
\label{sec:conclusions}

In this paper, we have presented a multi-scale study of growth of
submonolayer 2D Fe islands on the Mo(110) surface.  By using data from
{\it ab initio} electronic structure calculations, we obtained the
single-particle diffusion barrier $E_d$ from the adiabatic potential
surface seen by a single Fe adatom. The same calculations were used to
determine interactions between pairs of Fe atoms up to 6th nearest
neighbor pairs.  These were used to construct a simple lattice-gas
Hamiltonian in equation (1).  Such a Hamiltonian is immediately amenable to
direct KMC simulations of growth; however, to study time and
especially length scales relevant to MBE growth a coarse-grained
approach is beneficial. To this end we employed the KRE approach
within the framework of a simple aggregation model of islands. The
adatom island diffusion coefficients, which are needed as input to
KRE, were determined from equilibrium MC simulations using the
lattice-gas Hamiltonian, up to sizes of a few tens of atoms at $T=500$
and $1000$ K. These data were further used to simplify the
interactions by including them up to 3rd nearest neighbors only in the
Hamiltonian.

Following construction of the KMC and KRE models we then carried out
actual MBE growth simulations for both models at the saturation
coverage $\theta = 0.15$. Tests using the KMC data revealed that the
scaled distribution function of island sizes $g(x)$ is rather
insensitive to the flux in the regime between about $1$ and $100$
ML/s, and thus the latter value was used in most cases to facilitate
comparison between KMC and KRE data. At $T=500$ K such a comparison
revealed that the mean-field type KRE model gives very good agreement
in the full island-island mode with KMC data for large islands, but the agreement
is poor for the smallest
islands, where KRE overestimates the regime $x \ll 1$. Comparison between
island-island and adatom-island modes in the KRE model also reveals
that at this temperature including single adatom mobility only gives
reasonably good results for $g(x)$. On the other hand, the same
comparison at $T=1000$ K (where KMC simulations could not be performed
even with $100$ ML/s) shows that the scaled distribution function is
even qualitatively wrong if island mobility is not included in the KRE
aggregation rates.

\ack
The authors acknowledge the support from
the Grant Agency of the Czech Republic under Contract No. 202/09/0775,
and by joint funding under EU STRP 016447 MagDot and NSF DMR Award
No. 0502737. This work has also been supported in part by the Academy of Finland
through its COMP Centre of Excellence Program (project No. 251748).
Computer resources provided by CSC-Center for
Scientific Computing Ltd. are also
acknowledged.

\section*{References}
\bibliography{multiscale_FeMo}

\end{document}